# Controllable Planning, Responsibility, and Information in Automatic Driving Technology

Dan WAN, Han ZHAN[*]

Institute for AI Moral Decision-Making, Hunan Normal University, Changsha, China

zhanhao@smail.hunnu.edu.cn

**Abstract**     People hope automated driving technology should be always in a stable and controllable state, accurately, which can be divided into controllable planning, responsibility, and information. Otherwise, it would bring about the problems of tram dilemma, responsibility attribution, information leakage, and security. This article discusses these three types of issues separately and clarifies some misunderstandings.

## 1 Introduction

Automatic driving technology has become the focus of research institutions and manufacturers all around the world. Both traditional automakers and Internet companies have long been involved in the development of automatic driving technology and have achieved certain results. In 2017, GM equipped the Super Cruise automatic driving function on the Cadillac CT6. In April of the same year, Baidu released the Apollo self-driving car platform. In July, Audi officially released the Audi A8, and its automatic driving system Traffic Jam Pilot reached Level 3. In October, Waymo completed the first social road test of Level 4 self-driving cars for the first time. In April 2018, Baidu launched the test ride of Level 4 Baidu driverless bus "Apolon", and announced the automatic driving bus entered the mass production phase in July.[1]

The rapid development of automatic driving technology has also led to a lot of discussions - most of which are concerned about the widespread use of automatic driving technology. Some people believe that there are many ethical and legal dilemmas in automatic driving technology, and they must be constrained to meet people's needs before they are actually applied to real life.

It is hoped that automatic driving technology will always be in a controlled state while reducing the driving accident rate. This means that we do not impose strict restrictions on automatic driving technology. While recognizing its shortcomings, we effectively recognize and control the cost of using this technology, so that it is in a state of controllable balance.[2]

The controllability of automatic vehicles is reflected in three points:

---

[*] Corresponding author.

[1] The Level 2, Level 3 and Level 4 refer to the automatic driving rating system of SAE, see http://standards.sae.org/j3016_201401/

[2] No matter it is the attribution of Level 3 accident responsibility or the prevention of leakage of driving record data, it is the pursuit of controllability. It will be stated in detail in the following passage.

(1) Controllable Planning, and its representative problem is the trolley problem;

(2) Controllable Responsibility, and its representative problem is responsibility belongings;

(3) Controllable Information, and its representative problems are information leakage and information security.

And people's doubts about automatic driving technology are mainly concentrated in these three parts. From another point of view, as long as the planning, responsibility and information controllability problems of the automatic driving technology are realized, the doubts about the automatic driving technology are solved to some extent.

## 2 Controllable Planning: Trolley Problem Under the Automatic Driving Scene

Planning is the decision and command issued by the automatic driving system to the automatic vehicles, that is, the path planning and manipulation instructions, such as "what kind of driving path is legal" or "when faced with an inevitable accident, what kind of decision is ethical" are all questions about whether the automatic driving system can truly achieve " controllable planning ". Among them, the most typical martyrdom is the trolley problem.

Trolley problem was initially proposed by Philippa Foot (1967, pp.152–161): a driver driving a "runaway tram" on a forked track, with five people on the original route tied to the track. If the trolley continues to drive, then the five will die. And on the other road, there is only one person tied to the track. The driver is faced with two choices: either doing nothing so that the tram will hit five people; or turn to another road so that only one person has to be sacrificed. What would you do if you are a driver?

On this basis, Judith Jarvis Thomson presented a series of more detailed trolley problems (1976, pp.204-217). After that, there have been more variants of the trolley problem, involving whether the decision makers themselves are dead or not (Bonnefon 2016, pp.1573-1576). After the rise of automatic driving, the discussion of the trolley problem under automatic driving technology has also become hot (Bonnefon 2015; Awad, Edmond, et al., 2018, p.59; Shariff, Azim, et al. 2017).

However, people have neglected (1) the trolley problem defaults in any decision exist losses; (2) for the automatic system, if you want to make decisions under the tram dilemma, you need to set up a program to let it make corresponding decisions when faced with similar situations. The above two points mean that we need to make a loss of decision when designing the autopilot system. In other words, the automatic driving system will make decisions to sacrifice humanity in a certain situation, which is difficult for us to accept.

In fact, as early as 2016, Mercedes-Benz said in public that the automatic driving system may give priority to protecting the safety of passengers in the car under the premise of damage, and sacrifice the outside people, which was strongly resisted by the public. People think that the automatic driving system has no right to make life choices.[3] On the other hand, as the official guidance document for the ethical code of self-driving cars, the German Ethics

---

[3] See, http://www.xinhuanet.com//world/2016-10/16/c_129323922.htm

Commission on Automatic and Connected Driving also explicitly mentions that the autopilot system with "loss prediction" is not allowed.[4]

Once "loss prediction" is rejected into the automatic driving system, the tram dilemma under the automatic driving based on the lossy preset will no longer exist, and our doubts about the "controllable planning" will be resolved.

## 3 Controllable Responsibility: from Responsibility to Business Behavior

Responsibility attribution problem is another widely discussed issue in automatic driving technology: How should the liability of the accident be determined if an accident occurs during automatic driving?

In the automation grading system introduced by the Society of Automotive Engineers (SAE), automatic vehicles are divided into six levels: Level 0 No Automation; Level 1 Driver Assistance; Level 2 Partial Automation; Level 3 Conditional Automation; Level 4 High Automation; Level 5 Full Automation.[5] Level 0, Level 1, and Level 2 belong to the assisted driving phase, the automatic driving system only assists human driving, but cannot perform individual driving behavior. In Level 3, the system can independently complete most driving operations, but when the emergency occurs, the driver needs to take over. In Level 4, the system could complete all driving operations in some scenarios. In Level 5, the automatic car can complete the driving operation in any scene.

In L0, L1, and L2, the automatic driving technology only plays an auxiliary role, the division of responsibility is obviously same as traditional driving method, so there is no need to discuss the attribution of responsibility; in the L4, although all the automatic driving techniques have been implemented by automatic driving system, the whole process must be in a specific scene, such as freight passages and power plants, and the accident responsibility is easily defined clearly (SiXiao, CaoJianfeng 2017, pp.166-173). As for the L5, it is only an imaginary; the key in the responsibility allocation problem lies in L3.

Some researchers believe that the criminal responsibility should be excluded if we can prove the automated vehicles are causing damage within its automatic driving range, and the current technical level can't predict and prevent the damage situation (Long Min 2018, pp.78-83). Some scholars have pointed out that the responsibility of the company can be analyzed by judging the possibility of avoidance, the decision-making ability of the driving system, and the supervision responsibility of enterprises and personnel who are manufacturing, producing, and programming (Cheng Long 2018, pp.83-89).

However, these schemes are neither accepted by the public nor give actual solutions, so they are not able to properly resolve the responsibility of automatic technology in L3. From the perspective of supporting the development of automatic driving, we proposed two solutions: 1) put the responsibility on the drivers and 2) technology level spanning.

Option 1: put the responsibility on the drivers

---

[4] See, https://www.bmvi.de/SharedDocs/EN/publications/report-ethics-commission-automated-and-connected-driving.pdf?__blob=publicationFile

[5] See, http://standards.sae.org/j3016_201401/

This option aims at driver's freedom to choose and merchandise attributes of the self-driving car.[6] On the one hand, drivers choose their self-driving cars or automatic driving system by their free wills, they should be responsible for those choices, so we can attribute the responsibility of the automatic driving accident to the drivers (If there are only two responsibility subjects, driver and driverless car).[7]

Some people will argue that, will it lead to no one to buy a driverless car if we attribute the accident responsibility to the driver? In fact, it will not. Seller or manufacturer will make a responsible commitment agreement based on the consideration of cost and accident probability for the purpose of profit, then transfer the responsibility of the accident from the driver to themselves in a form of compensation. This transfer can be guaranteed not only in the form of a sales clause but also the insurance contract.

Option 2: technology level spanning

This option completely solves the responsibility problem of L3 by forbidding cross-driving between a self-driving car and human-driving car. In fact, Volvo, Ford, and other companies have claimed that they will abandon the automatic driving technology research and skip L3 to L4 or L5, because of the responsibility issues in L3. Similarly, many scholars proposed that companies should skip L3 in view of the liability problem of L3 on the Netease 2017 Future Technology Summit.[8]

## 4 Controllable Information: Leakage and Security

The last controllable obstacle to automatic driving is a controllable message. Specifically, they are the leakage problem and the security problem.

The first one means there is the danger of privacy violation and data leakage when automatic vehicles have to continuously collect relevant information and data (JiangSu 2018,pp.180-189). However, this kind of problem is not unique to automatic driving technology: mobile phone, computer, iPad, and other devices that people carry with can collect the driving data have the same trouble. For example, the APP "OKDrive" has realized to acquire the data of location, history, speed, rapid acceleration, number of sudden braking, etc. through the mobile phone GPS and sensor modules data.[9]

Therefore, there is not only the automated driving system`s privacy problem but an internet information`s privacy problem. So, it is not appropriate to question automatic technology by this.

Security problem mainly refers to the risk of attacked or maliciously invaded in automatic technology. Unlike leakage, security problem has its own particularities. The safety problem of traditional vehicles has the following characteristics: 1) physical contact; 2) the number of the car be damaged is limited in a short time; 3) easy to be detected. While for the self-driving car: 1) no physical contact; 2) multiple cars can be destroyed in a short time; 3) difficult to be detected.

---

[6] Assume that the purchaser of the automatic-driving vehicle is the driver himself.
[7] This is similar to the traditional driving scene where people close their eyes and step on the gas pedal, letting luck decide whether to cause a car accident. This act of fully entrusting security to "destiny" is similar to entrusting security to "automatic technology." Based on this consideration, we can attribute the responsibility of the accident to the driver.
[8] See, http://live.163.com/room/141395.html
[9] See, http://www.okchexian.com/okdriveEng.html

Fortunately, the technology community has recognized this problem and proposed many solutions. At present, we can solve this problem with intensive learning based on confrontation.[10]

## 5 Conclusion

For a variety of reasons, misunderstandings always exist in our understanding of autonomous driving systems. Indeed, there are many problems with the current autonomous driving system, and that is why we should clarify these misunderstandings and focus on real problems.

**References**

[1] Awad, Edmond, et al., 2018, "The moral machine experiment", Nature, 563.7729.
[2] Bonnefon, Shariff, Rahwan (2016), "The Social Dilemma of Autonomous Vehicles", Science, 352(6293).
[3] Bonnefon, Shariff, Rahwan (2015), "Autonomous vehicles need experimental ethics: Are we ready for utilitarian cars", arXiv preprint, arXiv:1510.03346.
[4] Foot (1967). "The problem of abortion and the doctrine of double effect", Oxford Review, 2(2).
[5] Shariff, Azim, Bonnefon, Rahwan (2017), "Psychological roadblocks to the adoption of self-driving vehicles", Nature Human Behavior, 1.10.
[6] Thomson, Jarvis (1976). Killing, letting die, and the trolley problem, The Monist, 59(2).
[7] JiangSu (2018), "The Challenges of Self-driving Cars to the Law", China Law Review, (2).
[8] SiXiao, CaoJianfeng (2017), "On the Civil Liability of Artificial Intelligence", Science of Law (Journal of Northwest University of Political Science and Law, 35(5).
[9] LongMin (2018), "Assignment of the Criminal Liability in Automatic Driving's Accidents", ECUPL Journal, 21(06).
[10] ChengLong(2018), "The Criminal Regulation of Traffic Accidents Caused by Autonomous Vehicles", Academic Exchange, No.289(04).

---

[10] See, https://arxiv.org/pdf/1803.00916.pdf